\pdfoutput=1
\documentclass[a4paper,10pt,amsmath,amssymb,nofootinbib,onecolumn,
floats,longbibliography,aps,prd]{revtex4-2}
\usepackage[british]{babel}

\usepackage{mathrsfs,bm}
\usepackage{natbib,textcase}
\usepackage{graphicx,color,soul,comment}
\usepackage[pdftex,pdfusetitle]{hyperref}

\hypersetup{colorlinks=true,linkcolor=red,citecolor=blue,filecolor=black,urlcolor=black, pdfauthor={Jacopo Mazza}}

\usepackage{graphicx,color,soul}
\usepackage[pdftex,pdfusetitle]{hyperref}

\usepackage[capitalize]{cleveref}
\Crefname{figure}{Fig.}{Figs}
\crefname{subequation}{Eqs.}{Eqs.}
\labelcrefformat{subequation}{#2(#1)#3}
\crefformat{subequations}{#2(#1)#3}
\crefname{section}{Sec.}{Secs.}
\Crefname{section}{Section}{Sections}

\allowdisplaybreaks

\usepackage[caption=false]{subfig}
\usepackage{physics}
\usepackage{array,booktabs} 
\usepackage{multirow} 
\newcolumntype{L}{>{$}p{20mm}<{$}} 
\makeatletter\g@addto@macro\bfseries{\boldmath}\makeatother%

\allowdisplaybreaks

\def\be#1\ee{\begin{align}#1\end{align}} 

\newcommand{\iu}{\mathrm{i}}

\usepackage{comment}

\begin{document}

\title{Stable rotating regular black holes}

\author{Edgardo Franzin} 
\author{Stefano Liberati} 
\author{Jacopo Mazza} 
\author{Vania Vellucci} 

\affiliation{SISSA, International School for Advanced Studies, via Bonomea 265, 34136 Trieste, Italy}
\affiliation{INFN, Sezione di Trieste, via Valerio 2, 34127 Trieste, Italy}
\affiliation{IFPU, Institute for Fundamental Physics of the Universe, via Beirut 2, 34014 Trieste, Italy}

\begin{abstract}
We present a rotating regular black hole whose inner horizon has zero surface gravity for any value of the spin parameter, and is therefore stable against mass inflation.
Our metric is built by combining two successful strategies for regularizing singularities, i.e.\ by replacing the mass parameter with a function of $r$ and by introducing a conformal factor.
The mass function controls the properties of the inner horizon, whose displacement away from the Kerr geometry inner horizon is quantified in terms of a parameter $e$; while the conformal factor regularizes the singularity in a way that is parametrized by the dimensionful quantity $b$. The resulting line element not only avoids the stability issues that are common to regular black hole models endowed with inner horizons, but is also free of problematic properties of the Kerr geometry, such as the existence of closed timelike curves. While the proposed metric has all the phenomenological relevant features of singular rotating black holes --- such as ergospheres, light ring and innermost stable circular orbit --- showing a remarkable similarity to a Kerr black hole in its exterior, it allows nonetheless sizable deviations, especially for large values of the spin parameter $a$. In this sense, the proposed rotating ``inner-degenerate'' regular black hole metric is not only amenable to further theoretical investigations but most of all can represent a viable geometry to contrast to the Kerr one in future phenomenological tests.   
\end{abstract}

\maketitle 

\section{Introduction}

There is nowadays compelling direct evidence of the existence of dark compact objects.
While general-relativistic black holes (BHs) --- characterized solely by their mass~$M$ and specific angular momentum~$a$ --- are favorite to describe these astrophysical objects, beneath their horizon they house a classically inevitable singularity.
It is widely believed, though, that quantum-gravity effects change the internal structure of BHs at some scale $\ell$ and cure the singularity issue.
Without specifying the actual theory responsible for the singularity regularization, there is only a handful of viable alternatives to classical BHs, i.e.\ regular black holes (RBHs), wormholes, black bounces, and quasi black holes~\cite{Carballo-Rubio:2018jzw,carballo-rubio_opening_2020,carballo-rubio_geodesically_2020}.

Here we focus on RBHs, i.e.\ geometries that are almost indistinguishable from a classical BH for a distant observer, yet with a smooth regular core. (For a classification of possible asymptotic geometries for the core in the static case see Ref.~\cite{bargueno_global_2020}.)
Clearly, these spacetimes are not vacuum solutions of general relativity (GR) and --- at most --- they could be vacuum solutions of a high energy completion of the theory. However, while they might require a transient quantum gravitational phase at the core of gravitational collapse to be realized, they are, by construction, classical, regular and smooth everywhere geometries.
So, it is reasonable to probe deviations from their GR counterparts by making use of the Einstein field equations (interpreted as the low energy Lagrangian of our effective field theory) and by analyzing the so obtained effective stress--energy tensor.
Generally, such an effective stress--energy tensor will sport localized violations of some energy conditions, which need not be associated to any physical matter field. More naturally, these violations can be interpreted as stemming from the higher order terms in the effective field theory gravitational action associated to the low energy limit of the (still unknown) quantum gravitational theory responsible for the singularity regularization.

For spacetimes which are everywhere well-defined and analytical, RBHs can have a non-singular core by sporting inside the outer horizon or a wormhole throat or an inner horizon~\cite{carballo-rubio_opening_2020,carballo-rubio_geodesically_2020}. Here we shall focus on the second class of regular objects.\footnote{See however Refs.~\cite{simpson_black-bounce_2019,Franzin:2021vnj,mazza_novel_2021} for static or rotating examples of the first class and Refs.~\cite{Franzin:2022iai, Riaz:2022rlx} for some of their phenomenological implications.}
The viability of such a class has been recently questioned by the realization that their inner horizon(s) are generically associated to an exponential instability~\cite{Carballo-Rubio:2018pmi,carballo-rubio_inner_2021,Carballo-Rubio:2022kad} known as mass inflation~\cite{Poisson:1989zz} regulated by their surface gravity, which is generically non-zero.\footnote{While it has been claimed that such an exponential instability is replaced by a milder polynomial growth in some RBH models~\cite{Bonanno:2020fgp}, it was later shown that the onset of such behavior is generically preceded by an exponential, and hence fatal, phase~\cite{carballo-rubio_inner_2021,DiFilippo:2022qkl}.} Remarkably, this instability has been proven to be avoidable in a recent spherically symmetric RBH proposal where a solution endowed with a degenerate inner horizon, characterized by a vanishing surface gravity, was constructed and shown to be stable~\cite{Carballo-Rubio:2022kad}. In what follows we shall generalize this idea and extend it to the phenomenologically more relevant rotating RBH spacetimes.

Indeed, the first RBH models have been introduced assuming spherical symmetry~\cite{ansoldi_spherical_2008}, which of course lacked a key feature for being astrophysical GR-black-hole mimickers: rotation. Over the years, there have been several proposals for rotating RBHs~\cite{Torres:2022twv}, often by means of the Janis--Newman procedure.
Even in this case, the required matter content to sustain the solution can be interpreted as associated to some non-linear electrodynamics~\cite{dymnikova_regular_2015,turimov_generic_2021,toshmatov_rotating_2014} and suitable criteria for reasonable RBHs were proposed so to rule out some models, or support others~\cite{maeda_quest_2021}.

Apart from the above-mentioned mass-inflation instability, an unpleasant common feature of the proposed rotating spacetimes is that the curvature invariants are found to be finite but not continuous at the would-be ring singularity~\cite{bambi_rotating_2013}. 
In addition, as for Kerr BHs, rotating RBHs generically still allow for closed timelike curves (CTCs).
Nonetheless, these models represented the simplest alternatives to Kerr BHs and for this reason their phenomenological features have been extensively studied~\cite{abdujabbarov_shadow_2016, Tsukamoto:2017fxq, ghosh_ergosphere_2020,Kumar:2020yem, akiyama_first_2022,kumar_walia_testing_2022, banerjee_signatures_2022}. In order to overcome the above-mentioned shortcomings of the so-far proposed RBH models, it is worth analyzing in more detail how these are normally constructed, i.e.\ how singularities are regularized.

The most common approach to model RBH geometries is by considering the Schwarzschild or Kerr metric and promoting its mass $M$ to be a function of the radial coordinate, $M\mapsto m(r)$. This is done in most of the above-mentioned proposals and generally amounts to requiring a $m(r)$ which goes to zero suitably fast close to the singularity and that after a transient region goes to the prescribed black-hole Arnowitt--Deser--Misner (ADM) mass $M$ at infinity.

Alternatively, it is also possible to remove the singularity by means of a suitable Weyl rescaling~\cite{narlikar_space-time_1977} of the metric, i.e.\ by multiplying the Schwarzschild or Kerr metric by a conformal factor~\cite{bambi_spacetime_2017,jusufi_rotating_2020}, or simply of some metric components~\cite{Canate:2022zst}.

In this paper we propose to combine these two approaches to obtain a rotating RBH which is free of all the issues discussed above.
In particular the conformal factor will be used to regularize the singularity, while the mass function will be chosen so to avoid the mass inflation along the idea proposed in Ref.~\cite{Carballo-Rubio:2022kad} (which cannot be as easily implemented in the rotating case by using only a mass function).
In summary, we shall show that our candidate spacetime has a number of desirable features: 
the curvature invariants are everywhere finite and continuous;
the region of the spacetime at $r<0$, responsible for CTCs, is inaccessible to timelike and null geodesics;
the inner horizon has zero surface gravity and as such is stable;
the outer/trapping horizon is located as in the Kerr spacetime;
the ergoregion, light ring and innermost stable circular orbit (ISCO) are similar, in shape and location, to their Kerr counterparts, yet allowing for, in principle, detectable differences.

\medskip

The paper is structured as follows. \Cref{sec:RBHs} introduces the basic ingredients for regularizing BHs, discusses some common pitfalls of traditional RBHs and introduces our strategy for avoiding them. \Cref{s:conformal} explains how a conformal factor can cure singularities and presents our own choice. \Cref{sec:stabilizing} shows how a carefully picked mass function can stabilize the inner horizon; we then recap the metric we are putting forward with \cref{eq:finalmetric}. In \cref{sec:mimicker} we investigate the extent to which the object we propose deviates from a Kerr BH by exploring some simple phenomenological applications: first we describe the casual structure, then the effective matter content, the shape of the ergosurfaces and the coordinate location of the light ring and ISCO\@. Finally, \cref{sec:conclusions} reports our conclusions.

\section[Basic ingredients of regular black holes]{Basic ingredients of regular black holes\label{sec:RBHs}}

Most static and spherically symmetric RBHs can be described by the line element
\be\label{eq:seed}
ds^2 = -\left(1-\frac{2m(r)}{r} \right) \dd{t}^2 + \left(1-\frac{2m(r)}{r} \right)^{-1}\dd{r}^2 + r^2 \left[ \dd{\theta}^2 + \sin^2 \theta \dd{\phi}^2 \right]\, .
\ee

For large $r$, $m(r)\to M$, i.e.\ to the ADM mass, and the spacetime is asymptotically flat.
Regularity is ensured if the mass function $m(r) = \order{r^3}$ as $r\to0$~\cite{frolov_notes_2016} and $\lim_{r\to 0^+} m'(r)/r^2 \gtreqqless 0$ corresponds respectively to a de Sitter, Minkowski and anti-de Sitter core.

Some widely studied choices of $m(r)$ are summarized in \cref{tab:RBHs}. For all of these RBH models, thanks to the introduction of some characteristic (supposedly quantum-gravity-induced) length scale $\ell$, one finds that $r=0$ is a (regular) ``point'' --- it is the degenerate sphere obtained by shrinking constant-$r$ surfaces.
An interesting alternative, which we will not explore in this article, consists in endowing the center with a non-zero size: popular examples are black bounces~\cite{simpson_black-bounce_2019,mazza_novel_2021,Franzin:2021vnj} and Bronnikov black universes~\cite{Bronnikov:2005, Bronnikov:2006}.

\begin{table}
\centering
\begin{tabular}{ll}
\toprule
Model & $m(r)$\\
\midrule
Bardeen~\cite{bardeen_non-singular_1968} & $M\frac{r^3}{(r^2+\ell^2)^{3/2}}$\\
\midrule
Hayward~\cite{hayward_formation_2006} & $M \frac{r^3}{r^3 + 2M\ell^2}$ \\
\midrule
Dymnikova~\cite{dymnikova_vacuum_1992} & $M \left[1-\exp(-\frac{r^3}{\ell^3}) \right]$\\
\midrule
Fan--Wang~\cite{Fan:2016hvf} & $M \frac{r^3}{(r+\ell)^3}$\\
\bottomrule
\end{tabular}
\caption{Some of the most popular RBH models. More examples e.g.\ in Refs.~\cite{Ghosh:2014pba,Tsukamoto:2017fxq,Kumar:2018ple,Kumar:2020yem,simpson_eye_2022}\label{tab:RBHs}}
\end{table}

Perhaps the most straight-forward way to add rotation to these spacetimes is to repeat the same ``regularization'' on the Kerr metric, replacing its mass parameter $M$ with the function $m(r)$. The resulting line element is known as G\"urses--G\"ursey metric~\cite{Gurses:1975vu}:
\be\label{metricGG}
ds_\text{GG}^2 = -\left(1-\frac{2m(r)r}{\Sigma}\right) \dd{t}^2 - \frac{4a \, m(r) r \sin^2 \theta }{\Sigma} \dd{t} \dd{\phi} + \frac{\Sigma}{\Delta} \dd{r}^2 + \Sigma \dd{\theta}^2 + \frac{A \sin^2 \theta}{\Sigma} \dd{\phi}^2
\ee
with
\be
\Sigma = r^2 + a^2\cos^2\theta, \quad \Delta=r^2-2m(r)r +a^2, \quad A= (r^2+a^2)^2 -\Delta a^2 \sin^2 \theta\, .
\ee

Usually, the same $m(r)$ is used in the rotating and non-rotating cases, although there might be reasons to introduce a dependence on the angle~\cite{eichhorn_locality-principle_2021, eichhorn_image_2021-1}.
In the G\"urses--G\"ursey spacetime, a necessary and sufficient condition to avoid scalar curvature singularities is $m(0)=m'(0)=m''(0)=0$~\cite{torres_regular_2017}.
Basic properties of the line element \eqref{metricGG} as a rotating RBH are discussed in Ref.~\cite{dymnikova_basic_2017}.

Killing horizons are located at values of $r$ that solve $\Delta = 0$; since this equation must have more than one root, these spacetimes generically possess inner horizons. As in the Kerr spacetime, such inner Killing horizons coincide with Cauchy horizons, hence these spacetimes are not globally hyperbolic.

Metrics of this form have been extensively used for phenomenological applications. From a theoretical point of view, however, these examples are not completely satisfying, for several reasons.

First of all, their inner horizons have non-zero surface gravity and for this reason they are prone to a phenomenon known as mass inflation. As a consequence of this phenomenon, any perturbation --- no matter how small --- will lead to an accumulation of energy close to the inner horizon and, eventually, to a large backreaction on the geometry. Since this phenomenon proceeds exponentially fast, it is regarded as a source of instability. Hence, RBHs as those described above can at most be metastable. 

Secondly, the interpretation of the region close to the center is far from clear. As we said, in the non-rotating case the metric approaches that of an (anti) de Sitter or Minkowski spacetime close to $r=0$. As rotation is included, however, the would-be singularity becomes a ring located at $r=0, \ \theta=\pi/2$ and the simple interpretation in terms of maximally symmetric spaces is lost. 
A particularly puzzling issue concerns the behavior of the curvature scalars close to the would-be singularity: the limits $r\to 0$ and $\theta \to \pi/2$ often do not commute (typically, if the $r$ limit is taken first the result is zero, but if the $\theta$ limit is taken first the result is a number proportional to inverse powers of~$\ell$).
Hence, strictly speaking, the limit does not exist.\footnote{The $\Sigma\to0$ limit of the curvature invariants is well defined, and zero, if the mass function goes to zero faster than $r^3$. This is not the case for the most popular RBH models, but it is so for e.g.\ a Bardeen-like RBH with $m(r)=r^4/(r^2+\ell^2)^2$, as remarked also in Ref.~\cite{Toshmatov:2017zpr}, and for the metric introduced in Ref.~\cite{simpson_eye_2022}.}

Moreover, these spacetimes usually admit an analytical extension to a region of negative radial coordinate, where other pathologies might still arise. In particular, the $r<0$ region is responsible for the existence of CTCs and of a chronological horizon in the Kerr spacetime: since there seems to be no general mechanism to prevent the formation of CTCs in these RBHs, their existence becomes a model-dependent question.  

\medskip

The line element \eqref{metricGG} can also be obtained via the Newman--Janis procedure with appropriate choices of complexification.
To avoid the intrinsic ambiguities of the standard implementation, Azreg-A\"inou~\cite{azreg-ainou_generating_2014, azreg-ainou_regular_2014, azreg-ainou_static_2014} introduced a modified version of the procedure that leads to a metric whose components are the same as in \cref{metricGG}, up to a free multiplicative function $\Psi(r, \theta)$,
\be\label{eq:PsiGG}
ds^2 = \frac{\Psi}{\Sigma}\,ds_\text{GG}^2\,,
\ee
where the free function $\Psi$ can then be constrained by some physical arguments. In particular, if the seed metric can be interpreted as a solution of some non-linear electrodynamics, the field equations imply $\Psi=\Sigma$.
Moreover, within the Newman--Janis framework, in the $a\to0$ limit, the line element \eqref{eq:PsiGG} must reduce to the seed metric, i.e.\ $\lim_{a\to 0} \Psi(r, \theta) = r^2$.
The choice $\Psi = \Sigma$ thus seems the most natural, but it entails all the issues we just discussed.

On the other hand, a large body of work in the context of conformal gravity has shown that appropriate choices for the conformal factor can lead to geodesically complete spacetimes. In light of this fact, \cref{eq:PsiGG} is particularly suggestive, as it encodes a significantly larger amount of freedom than \cref{metricGG}. In particular, when $\Psi/\Sigma \neq 1$ it becomes possible to disentangle the regularization of the central singularity from the choice of $m(r)$, which instead determines the location and properties of the horizons. 

The goal of this paper is therefore to show how one can exploit such freedom to build a RBH that is free of the issues presented above: in our proposal, $\Psi$ will be used to improve the appearances of the spacetime close to $r=0$, while $m(r)$ will be chosen so as to trim the properties of the inner horizon.

Note, incidentally, that adding a conformal factor to \cref{metricGG} is by no means disruptive: the spacetime described by the metric \eqref{eq:PsiGG} is of Petrov type D, exactly as \cref{metricGG}, with double null directions given by
\be
l^\mu = \sqrt{\frac{\Sigma}{\Psi}} \frac{1}{\Delta} \left\{r^2+a^2, \Delta, 0, a \right\}, \quad
n^\mu = \sqrt{\frac{\Sigma}{\Psi}} \frac{1}{\Delta} \left\{r^2+a^2, -\Delta, 0, a \right\};
\ee
these two null vectors can be complemented with
\be
m^\mu = \sqrt{\frac{\Sigma}{\Psi}} \frac{1}{\sqrt{2}\left(r + \iu a \cos \theta\right)} \left\{\iu a \sin \theta, 0, 1, \iu \csc\theta \right\} 
\ee
and its complex conjugate $\Bar{m}^\mu$ to form a Kinnersley-like tetrad. When  $\Psi(r, \theta)=\psi_r(r) + \psi_\theta (\theta)$ (i.e.\ it is ``separable''), the geometry admits the non-trivial Killing tensor
\be\label{eq:KT}
K_{\mu \nu} = \Psi(r, \theta) \left[l_\mu n_\nu + l_\nu n_\mu \right] + \psi_r (r) g_{\mu \nu}\, .
\ee
In this case, the equations of motion for a test particle of Killing energy per unit mass $E$ and Killing angular momentum along the axis of rotation per unit mass $L$ are
\be
\Dot{t} &= \frac{1}{\Psi\Delta}\left[A E - 2m(r) a r L \right] \label{tgeod} \,, \\
\Dot{\phi} &= \frac{1}{\Psi\Delta} \left[ \frac{L}{\sin^2\theta}\left(\Sigma-2m(r) r\right) +2m(r) a r E\right] \label{phigeod} \,, \\
\Psi^2 \Dot{r}^2 &= \left[(r^2 +a^2)E -aL\right]^2 - \Delta \left(\delta \psi_r +K \right)  \label{rgeod} \,, \\
\Psi^2 \Dot{\theta}^2 &= K-\delta \psi_\theta - \left(aE\sin\theta -\frac{L}{\sin\theta} \right)^2 = Q + \cos^2\theta \left(E^2a^2 - \frac{L^2}{\sin^2\theta} \right) - \delta \psi_\theta \label{thetageod} \,,
\ee
where $\delta=0$ for massless particles and $\delta=1$ for massive ones, while $K$ is the conserved quantity associated to the Killing tensor \eqref{eq:KT}, $K=K_{\mu \nu}\Dot{x}^\mu \Dot{x}^\nu$, and $Q = K -(Ea-L)^2$. Clearly, planar equatorial orbits are possible only if $\psi_\theta(\pi/2)=0$.  

In the more general case in which $\Psi$ is not separable, the equations of motion are more involved and not separable. Motion with $\Ddot{\theta} = \Dot{\theta}=0$ can take place on the equator and on the axis of symmetry if $\partial_\theta \Psi= 0$ there. Note that if $\Psi$ is a function of $\Sigma$ only this is always the case, since $\partial_\theta \Psi = \Psi' \partial_\theta \Sigma = 2a^2 \cos\theta \sin\theta \Psi'$.

\section{Regularizing the singularity with \texorpdfstring{$\Psi$}{Psi}}\label{s:conformal}

In this section we discuss how the function $\Psi$ can regularize the spacetime, regardless of the specific choice of $m(r)$. We assume such $\Psi$ will satisfy a very minimal set of requirements, namely: $\Psi(r,\theta)>0$ everywhere, in order to ensure that no additional singularities are introduced; and
\be
\frac{\Psi}{\Sigma} = 1+\order{\frac{1}{r^2}} \qq{as} r\to \infty\, ,
\ee
so that the spacetime ADM mass and specific angular momentum are still given by the parameters $M =\lim_{r\to \infty}m(r)$ and $a$, respectively --- this is tantamount to a slightly stricter version of the usual asymptotic-flatness condition. (Note, in particular, that we do not follow the physical interpretation of Refs.~\cite{azreg-ainou_generating_2014, azreg-ainou_regular_2014,azreg-ainou_static_2014} and hence we do not impose the partial differential equations that descend from that reasoning.)

Let us now look for a $\Psi$ that regularizes the singularity of the Kerr BH, i.e.\ one for which
\be
ds^2 = \frac{\Psi}{\Sigma}\,ds^2_\text{Kerr}
\ee
is the line element of a spacetime free of scalar polynomial curvature singularities. The same $\Psi$ will also regularize more general metrics characterized by a generic (analytic) $m(r)$. As will become clear momentarily, the function $\Psi$ can also ``remove'' regions of the spacetime with undesirable features.

A simple example of such $\Psi$ is
\be\label{eq:Psi}
\Psi = \Sigma + \frac{b}{r^{2 z}}\, ,
\ee
with $z$ a real number, which we will further constrain in a moment, and $b$ a positive constant with dimensions $[M]^{2z+2}$. Note that if $b \to 0$ as $M\to 0$ or as $a \to 0$ one may recover respectively the Minkowski or the Schwarzschild metric. The Ricci scalar has the form
\be
R = -\frac{6b\, r^{2 z}}{r^2\Sigma^2 \left(b+r^{2z} \Sigma \right)^3}P_z\left(r, \cos\theta\right)\, ,
\ee
where $P_z\left(r,\cos\theta\right)$ is an expression (a polynomial in $r$ and $\cos\theta$ when $z$ is an integer) that goes to zero at least as fast as $\Sigma^2$ in the limit $r\to 0,\ \theta\to \pi/2$.
Hence, the Ricci scalar never blows up for $z \geq 1$. However, $z=1$ still does not yield a well-defined limit, while for $z>1$ the limit exists and is zero, irrespective of the path taken to reach the would-be singularity in the $r$--$\theta$ space. Similar remarks hold for the Ricci tensor squared $R^{\mu \nu}R_{\mu \nu}$ and the Kretschmann scalar $R^\mu_{\ \nu \rho \sigma}R_\mu ^{\ \nu \rho \sigma}$. 

The ansatz in \cref{eq:Psi} can be written as $\psi_r(r) + \psi_\theta (\theta)$, i.e.\ it is ``separable'' in the terminology introduced at the end of \cref{sec:RBHs}, and hence has the advantage of leading to separable equations of motion.

Note that, with $z>1$, $\Psi$ is divergent on the whole disk $r=0$ --- which will have consequences for CTCs. The fact that this divergence can in fact cancel the divergences in the curvature scalars is quite remarkable. For these reasons, \cref{eq:Psi} is the choice we will mostly explore in the remainder of the paper: in particular, we will often consider the ``minimal'' choice $z=3/2$, corresponding to the smallest integer exponent of $r$ that yields a well defined limit.

It is also worth mentioning that, if one focuses on the non-spinning case only, lower values of the exponent $z$ are required.
Indeed, in order to regularize the metric
\be
ds^2 = \frac{\Psi}{r^2}\,ds^2_\text{Schw}\,,
\ee
with the $a\to 0$ limit of \cref{eq:Psi} 
\be
\Psi = r^2 + \frac{b}{r^{2z}}\,,
\ee
one must have $z \geq 1/2$.

Finally, before moving on, let us add that an interesting alternative to \cref{eq:Psi} can be represented by the ansatz
\be\label{eq:Psi2}
\Psi = \Sigma + \frac{b}{\Sigma^z} \,.
\ee
In this case it is easy to check that the Ricci scalar tends to zero for $r\to 0, \ \theta \to \pi/2$ for any $z>1$. The same holds true for the Ricci tensor squared and the Kretschmann scalar, hence \cref{eq:Psi2} seems equivalent to \cref{eq:Psi}. Notably, however, in this second case $z=1$ too yields a well-defined, and finite, limit
\be
\lim_{\Sigma \to 0} R = - \frac{24 a^2}{b}
\ee
and similar results can be found for $R^{\mu \nu}R_{\mu \nu}$ and the Kretschmann.\footnote{To our knowledge, those built with a $\Psi$ are the only examples of rotating RBHs whose curvature scalars are continuous and non-zero at the would-be singularity.}
With this choice, $\Psi$ only diverges on the ring $r=0, \ \theta=\pi/2$, but not on the disk $r=0, \ \theta\neq \pi/2$. \Cref{eq:Psi2} will be juxtaposed to \cref{eq:Psi} in \cref{subsec:r0} to highlight the properties that make us prefer the latter.

\subsection{The spacetime close to \texorpdfstring{$r=0$}{r=0} \label{subsec:r0}}

Although the scalar curvatures we computed are everywhere finite, the components of the metric still diverge for $\Sigma = 0$.
Previous works~\cite{bambi_spacetime_2017} have argued that the resulting spacetime is in fact geodesically complete, since the would-be singularity is reached in infinite proper time. Since our choice of conformal factor is slightly different from that discussed in Ref.~\cite{bambi_spacetime_2017}, we sketch the relevant computations below.

Consider first a particle moving on the equatorial plane $\theta=\pi/2$ and falling radially towards $r=0$. With $E$ and $L$ being the particle energy and angular momentum per unit mass, its radial velocity satisfies
\be
\Psi^2 \Dot{r}^2 = \mathcal{R}_\text{Kerr} - \delta \Delta (\Psi-r^2)\, .
\ee
Here $\delta = 0$ or $1$ for massless or massive particles respectively, and $\mathcal{R}_\text{Kerr}$ is the right-hand side of \cref{rgeod} with $\Psi=\Sigma$. The proper time it takes for the particle to fall from $r_0$ to $r$ is 
\be
\Delta \tau = - \int^r_{r_0} \frac{\Psi}{\left[\mathcal{R}_\text{Kerr} - \delta \Delta (\Psi-\tilde r^2) \right]^{1/2}} \dd{\tilde r}\, .
\ee
For both our ansatz (\cref{eq:Psi} or \cref{eq:Psi2}), one finds that on the equatorial plane $\Psi - r^2 = b\, r^{-2z} > 0$ ; therefore the infall time for massless particles ($\delta=0$) is shorter than that for massive particles ($\delta =1$).
(Obviously, this is true as long as $\mathcal{R}_\text{Kerr} - \delta \Delta (\Psi- r^2)>0$, i.e.\ only where the trajectory is classically allowed: where the condition is not met, such motion could not take place.)

Let us then focus on massless particles. 
At $r=0$, $\mathcal{R}_\text{Kerr}=a^2(Ea-L)^2$, while $\Psi$ diverges at least as fast as $1/r^2$.
We conclude that massless particles reach the would-be singularity in an infinite amount of proper time. Given the inequality above, the conclusion remains true for massive particles.

Next, consider a particle that falls along the axis of symmetry $\theta=0$.
Such particle could reach the disk $r=0$ without encountering the would-be singularity, and potentially cross it through its center. On-axis motion requires $L=0$, so the radial velocity now satisfies
\be
\Psi^2 \Dot{r}^2= -\delta \Psi \Delta + E^2(r^2+a^2)^2
\ee
where $\Psi$ is now evaluated at $\theta=0$. The infall proper time becomes in this case
\be
\Delta \tau = -\int_{r_0}^r \frac{\Psi}{\sqrt{(\tilde r^2 + a^2)^2E^2-\delta \Delta \Psi}} \dd{\tilde r}\, .
\ee
First of all, we can see that it is still true that massless particles fall in a shorter time than massive ones, therefore we again focus on the former. We have
\be
E \Delta \tau_\text{light} = -\int \frac{\Psi}{r^2+a^2} \dd{r}\, 
\ee
and with our ansatz we have
\be
\frac{\Psi}{r^2+a^2} = 1+b \begin{cases} (r^2+a^2)^{-1}r^{-2z} &\qq{for \cref{eq:Psi}}\\
(r^2+a^2)^{-(z+1)} &\qq{for \cref{eq:Psi2}.}
\end{cases}
\ee
In the first case, the integrand diverges faster than $r^{-2}$ as $r\to 0$, hence the particle will reach the would-be singularity in an infinite time. In the second case, instead, the integrand is everywhere finite. For massive particles, one can show that the infall time remains finite in the second case but, according to the inequality above, it is infinite in the first. 

Therefore, the two choices of $\Psi$ lead to a very different structure of the region close to the would-be singularity: in the first case, the whole disk $r=0$ is (regularized and) ``sent to infinity''; in the second case, only the ring $r=0,\ \theta=\pi/2$ is pushed away, so that particles can still cross the disk inside the ring. This is a non-negligible difference as in the case of ansatz \cref{eq:Psi} we end up precluding, to light or matter, access to that region of the Kerr geometry ($r<0$) characterized by the presence of CTCs. 

Let us stress that while usually such a region is taken to be nonphysical in the Kerr geometry, due to the fact that it is shielded by a Cauchy horizon which is widely (albeit non-unanimously) considered unstable, the same region would represent a problem for us once we shall have proceeded to stabilize the RBH inner horizon by making it degenerate.
It is henceforth even more pressing for a stable RBH to chose an ansatz such as \cref{eq:Psi} over that of \cref{eq:Psi2}.

\section{Stabilizing the inner horizon with \texorpdfstring{$m(r)$}{m(r)}}\label{sec:stabilizing}

In Ref.~\cite{Carballo-Rubio:2022kad} it was shown that the mass-inflation instability can be turned off if the surface gravity of the inner horizon $\kappa_{-}$ vanishes thanks to a wise choice of the mass function. The problem to extend this idea to the rotating case consists in the fact that adding this condition (i.e.\ having a degenerate inner horizon), in addition to the conditions necessary to remove the ring singularity, avoid CTCs as well as have a well defined limit to the regularized region, makes the task daunting if not impossible. While we do not have a no-go theorem in this sense, it is rather clear to a first investigation that, even if viable, such regular metrics would be too cumbersome for any application towards phenomenology.

We shall then pursue a different path here, starting from the realization that if we regularize the singularity with the conformal factor as above, the functional form of the mass function is left with very few constraints (namely it must be everywhere finite and it must reduce to the ADM mass $M$ at infinity), and can be easily shaped so to stabilize the inner horizon.

Since the surface gravity of the inner horizon $r_-$ depends on $m(r)$ as
\be
\kappa_{-} \propto \partial_r \Delta |_{r=r_-}\, ,
\ee
if we assume a rational-function form for the mass function, to have a vanishing $\kappa_-$ the inner horizon must be a degenerate root of $\Delta$
\be
\Delta \equiv r^2-2 m(r) r+ a^2 = 0 \implies (r-r_+)(r-r_-)^d = 0\, ,
\label{delta}
\ee
for some $d \in\mathbb{N}_{\geq 2}$. $d=2$ is not viable, since it implies that $m(r)$ has a pole at some positive $r$. Thus the minimal choice ends up being $d=3$ which implies (given also the required asymptotic behavior) a mass function of the form
\be\label{m}
m(r) = M\,\frac{r^2+\alpha r + \beta}{r^2 + \gamma r+ \mu}.
\ee

From \cref{delta}, it can been shown that $\beta$ cannot be zero and thus the limit of $m(r)$ for $r \to 0$ is not zero but the finite value $M \beta/\mu$.
In this form, $m(r)$ is parametrized by four coefficients, two of dimension $[M]$ ($\alpha$ and $\gamma$) and two of dimension $[M]^2$ ($\beta$ and $\mu$).
However, through \cref{delta}, they can all be expressed as functions of the position of the two horizons
\be
\alpha &=\frac{a^4+r_-^3 r_+ - 3 a^2 r_- (r_- + r_+)}{2 a^2 M}\,,\\
\beta &=\frac{a^2 (2 M - 3 r_- - r_+) + r_-^2 (r_- + 3 r_+) }{2 M}\,,\\
\gamma &= 2M - 3r_- - r_+\,,\\
\mu &=\frac{r_-^3 r_+}{a^2}\,.
\ee
If we choose $r_+=M+\sqrt{M^2-a^2}$, i.e.\ the outer horizon to coincide with its Kerr analog, our family of metrics can be parametrized in terms of $r_-$ only.
It is quite remarkable, and very relevant for phenomenological studies, that in spite of being located beyond a trapping horizon, the position of the inner horizon can matter for observables in the outside geometry. An example of this can be exposed by looking at the large-$r$ behavior of the mass function:
\be\label{masympt}
m(r) \sim M +\frac{M (\alpha -\gamma)}{r} + \order{1/r^2}\,,
\quad (r\to\infty)\,.
\ee
The second term in the above expansion could be interpreted as an electric charge, and could lead to a different quadrupole moment with respect to a Kerr BH.

The choice $\alpha=\gamma$ must be discarded as it forces the inner horizon to coincide with the Kerr one, and in turn implies a non-zero inner horizon surface gravity (actually the usual one for the Kerr geometry) making the conformal Kerr metric still unstable to mass inflation.

Nonetheless, we can introduce a parameter controlling the difference between the inner-horizon position in our geometry and in Kerr. This parameter will in turn control the difference $\alpha-\gamma$. Let us write then
\be
r_-\equiv\frac{a^2}{M+(1-e)\sqrt{M^2-a^2}}\,,
\label{eq:epar}
\ee
with $e\neq0$ and $e<2$ in order to ensure $0<r_-<r_+$. Further requiring the mass function to have no poles implies
\be\label{ebounds}
-3 - \frac{3M}{\sqrt{M^2 - a^2}} < e < 2\,,
\ee
where in the positive (negative) part of the interval $r_-$ is larger (smaller) than the Kerr inner horizon.

With the above choice, it follows that $\alpha-\gamma = \order{e^3}$ --- the same holds true for all the other coefficients in the large-$r$ expansion. This suggests that sizable deviations of $r_-$ from its Kerr value could translate into measurable differences in the value of the quadrupole moment, or in the periastron precession and the orbital frequency in a binary system~\cite{Ryan:1995wh}. Such differences would all be $\order{e^3}$, which entails that values of $\abs{e}$ close to one or smaller might be phenomenologically favored; but the possible impact of $e$ on astrophysical observables certainly deserves further scrutiny, which we leave for the future.

Let us also note that, with the parametrization \eqref{eq:epar}, the mass function becomes $m(r) = M + \order{e^3}$ and in particular $m(r_+)=M$. This entails, among other things, that the outer-horizon angular velocity is the same as in Kerr, while its surface gravity is
\be
\kappa_+ = \frac{\partial_r \Delta(r_+)}{2(r_+^2+a^2)} = \kappa_+^\text{Kerr} + \order{e^3}\, .
\ee
Moreover, $e\to2$ is an extremal limit similar to $a\to M$, since in this limit $r_-\to r_+$ and $\kappa_+\to 0$.

Of course, different choices from \cref{eq:epar} for $r_-$ are in principle possible but they are strongly limited by a series of sanity requirements:
the inner horizon must lie within the outer horizon for all values of $a$; 
$m(r)$ must go to $M$ asymptotically; 
the denominator of $m(r)$ must have no zeros (for all $r>0$), that is $\gamma^2<4\mu$;
all the coefficients of $m(r)$ must be finite for all values of $a$; 
the extremal limit $a \to M$ should remain thermodynamically unattainable and thus also the surface gravity of $r_+$ should become zero in this limit --- indeed this is possible only if $r_- \to r_+$ for $a \to M$.

\medskip

In conclusion, the complete form of our rotating ``inner-degenerate'' metric is
\be\label{eq:finalmetric}
ds^2=\frac{\Psi}{\Sigma} \left[-\left(1-\frac{2m(r)r}{\Sigma}\right) \dd{t}^2 - \frac{4a \, m(r) r \sin^2 \theta }{\Sigma} \dd{t} \dd{\phi} + \frac{\Sigma}{\Delta} \dd{r}^2 + \Sigma \dd{\theta}^2 + \frac{A \sin^2 \theta}{\Sigma} \dd{\phi}^2 \right],
\ee
with $m(r)$ given in \cref{m} and
\be
\Psi=\Sigma+\frac{b}{r^3}, \quad \Sigma = r^2 + a^2\cos^2\theta, \quad \Delta=r^2-2m(r)r +a^2, \quad A= (r^2+a^2)^2 -\Delta a^2 \sin^2 \theta\,,
\ee 
where for the power law of $\Psi$ we have chosen the lowest integer that makes the curvature scalars continuous and finite (see \cref{s:conformal}).

Fixing $r_+=r_+^\text{Kerr}$ and choosing $r_-$ as in \cref{eq:epar}, this metric represents a family of stable, rotating, CTC-free, regular spacetimes with two free parameters (beyond the usual spin one): the ``Kerr-deviation parameter'' $e$  and the ``regularization parameter" $b$.
Notice that for $a \to M$ the metric becomes conformal to the extremal Kerr, while for $a \to 0$ the metric becomes conformal to Schwarzschild.

\section{The rotating ``inner-degenerate'' RBH as a Kerr black hole mimicker\label{sec:mimicker}}

In this section we investigate the extent to which our metric \eqref{eq:finalmetric} can mimic a Kerr BH: first we describe the causal structure; then the effective matter content; the position of ergosurfaces; and finally the location of the light rings and the ISCOs.

\subsection{Causal structure}

\begin{figure*}[t]
\centering
\includegraphics[width=0.7\textwidth]{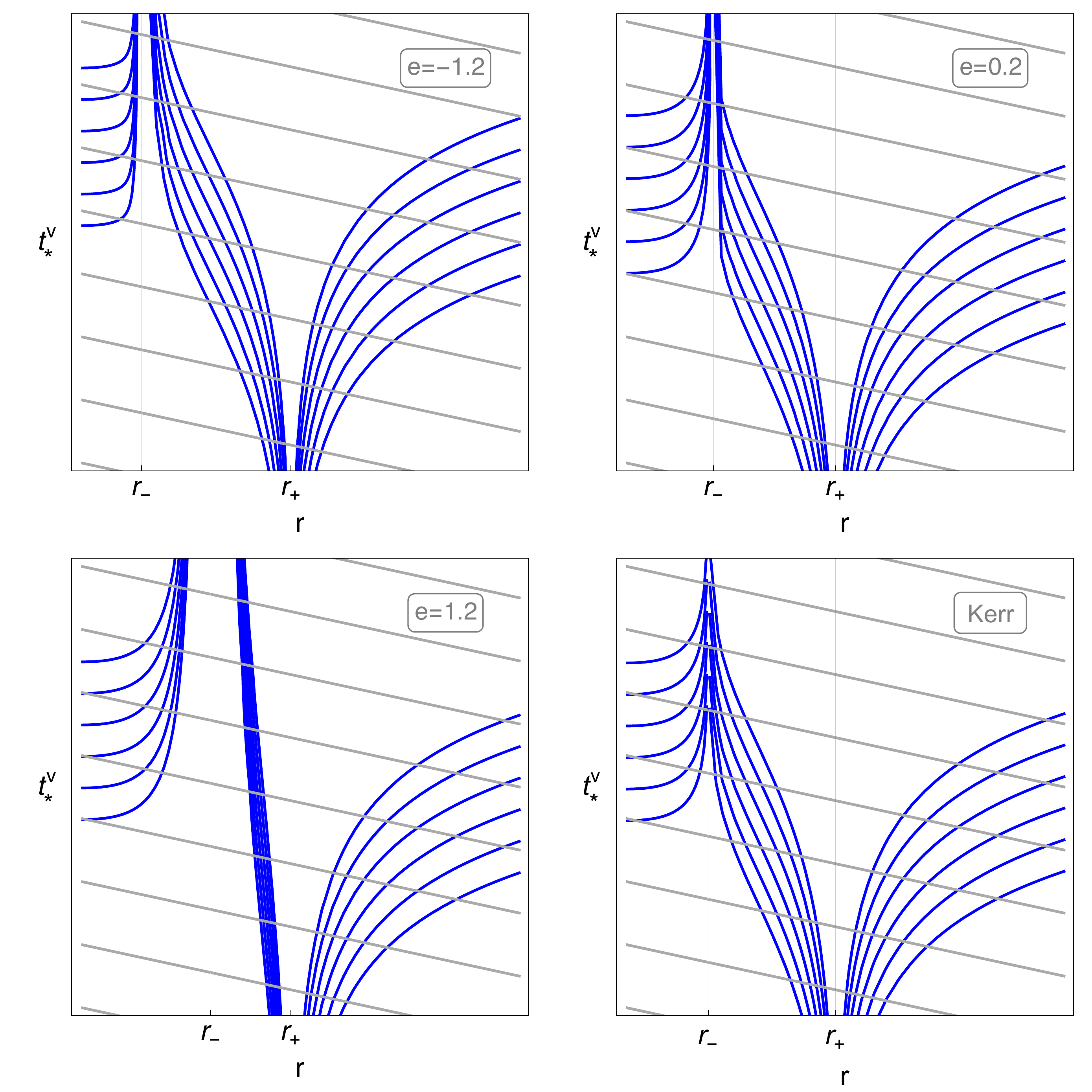}
\caption{Ingoing (gray) and outgoing (blue) null rays near the horizons  for selected values of the ``Kerr-deviation parameter'' $e$ compared with the Kerr ones in the bottom right panel. The spin parameter is set to $a=0.9M$.}
\label{fig:peel}
\end{figure*}

To study the casual structure of this spacetime we introduce ingoing null coordinates
\be
\dd{v} = \dd{t} + \frac{r^2+a^2}{\Delta} \,\dd{r}, \quad{   } \dd{\psi} = \dd{\phi} + \frac{a}{\Delta}\,\dd{r}\,,
\ee
that are regular at the horizons.
In \cref{fig:peel} we plot the equatorial principal null geodesics in the $r$--$t_*^v$ plane where $t_*^v$ is defined as 
\be
\dd{t}_*^v = \dd{v}-\dd{r}\,.
\ee
We see that, even if the inner horizon has zero surface gravity, we still have peeling of geodesics there, the difference with respect to Kerr is in the peeling trend.
Since $\kappa_- \propto \partial_r \Delta|_{r_-}=0$ and $\partial^2_r \Delta|_{r_-}=0$ this peeling is no longer exponential but scales as $1/t^{1/2}$.
In fact for the principal null geodesics
\be
\frac{\dd{r}}{\dd{t}}=\pm \frac{\Delta}{r^2+a^2}\,.
\ee
Thus near the inner horizon
\be
\frac{\dd{r}}{\dd{t}} &= 
\pm \frac{\partial^3_r \Delta|_{r_-}}{r_-^2+a^2}\,(r-r_-)^3 + \order{r-r_-}^4\,.
\ee

\begin{figure*}[ht]
\centering
\includegraphics[width=.65\textwidth]{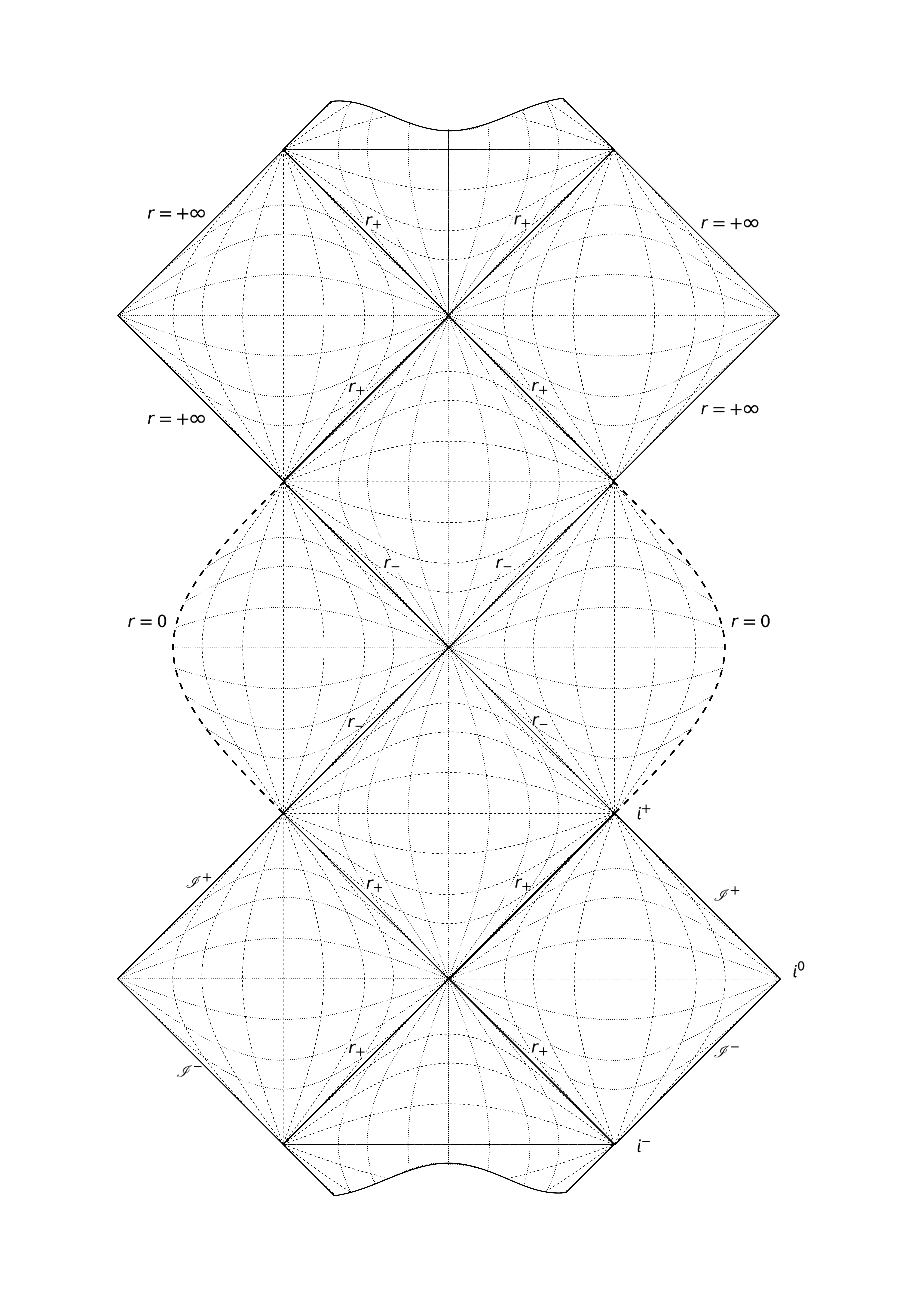}
\caption{Penrose diagram of the rotating RBH in \cref{eq:finalmetric}. The hypersurface $r=0$ is timelike, but reached in infinite proper time by any infalling observer: for this reason it is drawn not as a straight line but as a curve.}
\label{fig:penrose}
\end{figure*}

The causal structure of the spacetime is summarized by the Penrose diagram of  \cref{fig:penrose}. The diagram is completely analogous to that of the Kerr spacetime, except for the fact that the surface $r=0$ --- which is timelike --- is not a singularity and it can be reached only after an infinite amount of proper time by any infalling observer. In order to hint at these differences, we choose to represent $r=0$ as a branch of hyperbola instead of a straight line.

\subsection{Effective matter content}

Clearly, the metric we are considering is not a vacuum solution of GR\@.
Yet, as discussed in the Introduction, the Einstein equations can be used to characterize the spacetime by interpreting the Einstein tensor $G^\mu_{\ \nu}= R^\mu_{\ \nu}-\frac{1}{2}R\, \delta^\mu_{\ \nu}$ as an effective stress--energy tensor and to quantify deviations of our candidate spacetime with respect to the Kerr one.
To properly characterize the effective matter content, one first needs to project the Einstein tensor onto an orthonormal tetrad, e.g.\ the one of Refs.~\cite{azreg-ainou_generating_2014, azreg-ainou_regular_2014, azreg-ainou_static_2014}.
The behavior of the orthonormal components close to spatial infinity is particularly relevant: since the spacetime is asymptotically flat, they must all tend to zero as $r\to \infty$, but they do so at different rates. In particular, the slowest decaying (non-zero) components are those on the diagonal, all the others being of higher order in powers of $1/r$. Such components, at infinity, are the effective energy density and pressures:\footnote{Technically, the energy density and pressures are defined in terms of the eigenvalues of the orthonormalized Einstein tensor, when these are real. In asymptotically flat spacetimes, this procedure and the one presented in the text agree at leading order.}
\be\label{eq:rhop}
\rho = -p_r = p_\theta = p_\phi = -\frac{2M(\alpha-\gamma)}{r^4} +\order{1/r^5}\, .
\ee
Note that these quantities fall off quickly as $r\to \infty$, meaning that quantum-gravity-induced deviations from the GR vacuum solution are sizable only in a region close to the object. Moreover, they are $\order{e^3}$ and do not depend on $b$; the next-to-leading order $\order{1/r^5}$ also does not depend on $b$.
\Cref{eq:rhop} can lead to violations of the null energy condition (NEC), which requires $\rho + p_i \geq 0$, if $\alpha-\gamma >0$. When the NEC is violated, all the other classical energy conditions are violated too. When instead $\gamma > \alpha$, not only the null but also the weak (NEC + $\rho \geq 0$) and dominant ($\rho\geq \abs{p_i}$) energy conditions are met; the strong energy condition (NEC + $\rho+3p_i\geq 0$) instead is always violated.
Notice also that the above effective matter distribution does not correspond to any simple realistic matter content.
This is not surprising, as this effective stress--energy tensor gives an insight to the higher-order terms in the still unknown gravitational action.

Moving closer to $r=0$, the simple interpretation in terms of energy density and pressures is not always viable, since there are regions in which the Einstein tensor cannot be diagonalized over the real numbers: in these regions, the effective matter content is of type~IV in the Hawking--Ellis classification~\cite{hawking_large_1973}. (The existence of these regions is entirely due to the presence of the conformal factor: when $\Psi=\Sigma$, the effective stress--energy tensor is of Hawking--Ellis type~I for any $m(r)$.)

In order to circumvent this problem, we select particular geodesics and investigate the effective matter content as measured along them. 
We focus first on null geodesics: calling $k^\mu$ their tangent vector, the contraction
\be
G_{\mu \nu}k^\mu k^\nu
\ee
is always real and can be interpreted as the energy density measured along the geodesic. When this quantity is non-positive, the null energy condition is violated. 
For simplicity, we choose a geodesic that lies on the equatorial plane ($k^\theta =0$) and that falls towards the BH with zero angular momentum ($L=0$) --- cf.\ \cref{tgeod,phigeod,rgeod,thetageod}. Clearly, this choice represents a loss of generality, but is sufficiently illustrative for our purposes.

\begin{figure}[t]
    \centering
    \subfloat[\label{fig:rhoNplot1}]{\includegraphics[width=0.48\textwidth]{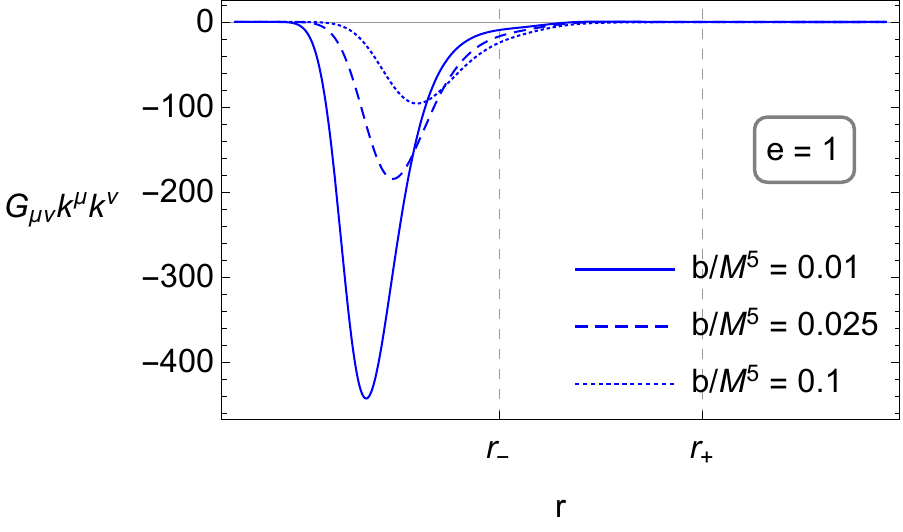}}
    \hfill
    \subfloat[\label{fig:rhoNplot25}]{\includegraphics[width=0.48\textwidth]{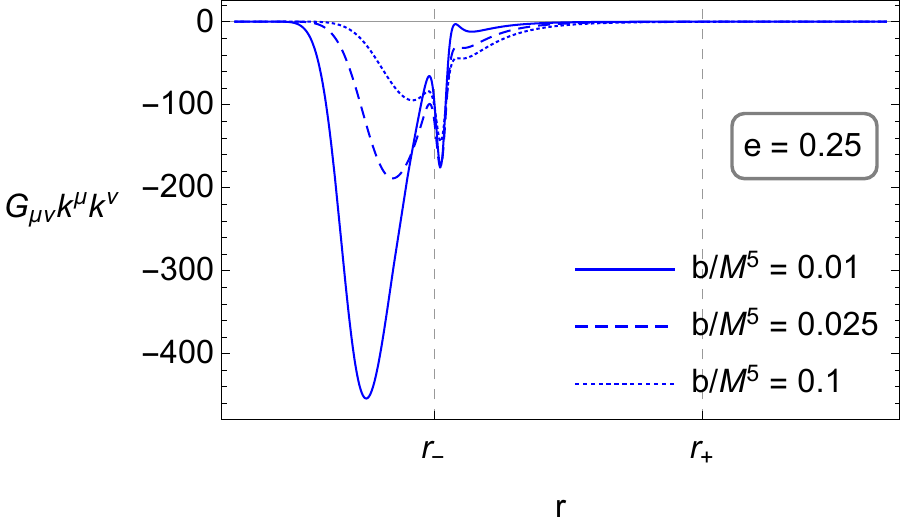}}
    \caption{Effective energy density as measured along a null equatorial trajectory, with $L=0$, that falls into a RBH with spin $a/M=0.9$. Each plot is relative to a particular choice of the deviation parameter $e$ and displays curves corresponding to three values of the regularization parameter $b$. The two vertical lines mark the location of the inner and outer horizons.}
    \label{fig:rhoNplot}
\end{figure}

The result is displayed in \cref{fig:rhoNplot}, for $a/M=0.9$ and some choices of the parameters $e$ and $b$. The effective energy density measured along the null geodesic is mostly negligible outside of the BH; inside the outer horizon, it becomes large and negative, signaling a substantial violation of the null energy condition; and it is exactly zero at $r=0$ (although that point is reached only at infinite affine parameter). The plot of \cref{fig:rhoNplot1} is representative of all the cases $\abs{e}\gtrsim 1$: increasing $e$ slightly moves the negative trough to the right; increasing $b$, instead, tends to smooth out the trough; but the overall shape of the curve is not greatly affected. When $\abs{e}\lesssim 1$, the curves exhibit additional features close to the inner horizon, signaling that the limit $e\to 0$ is not smooth. Lowering the spin suppresses the height of all the features just described.

\begin{figure}[t]
\centering
    \subfloat[\label{fig:SECMplot1}]{\includegraphics[width=0.48\textwidth]{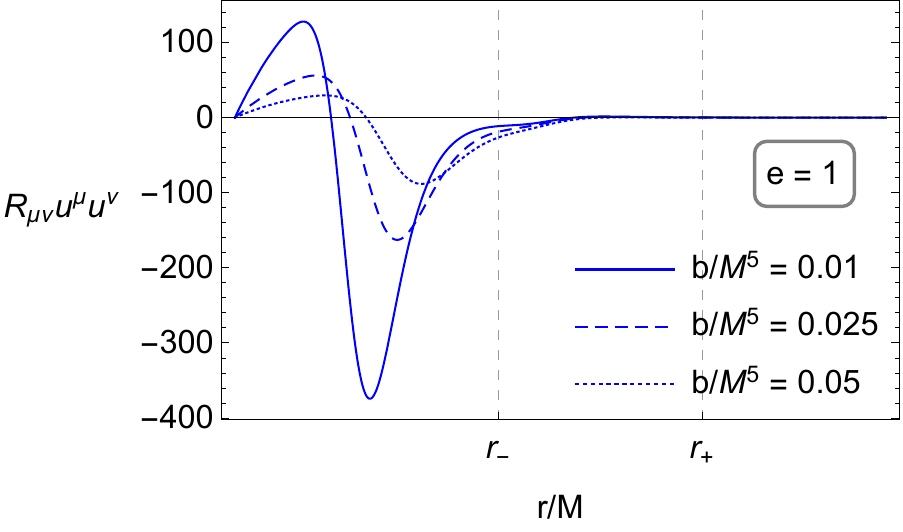}}
    \hfill
    \subfloat[\label{fig:SECMplot25}]{\includegraphics[width=0.48\textwidth]{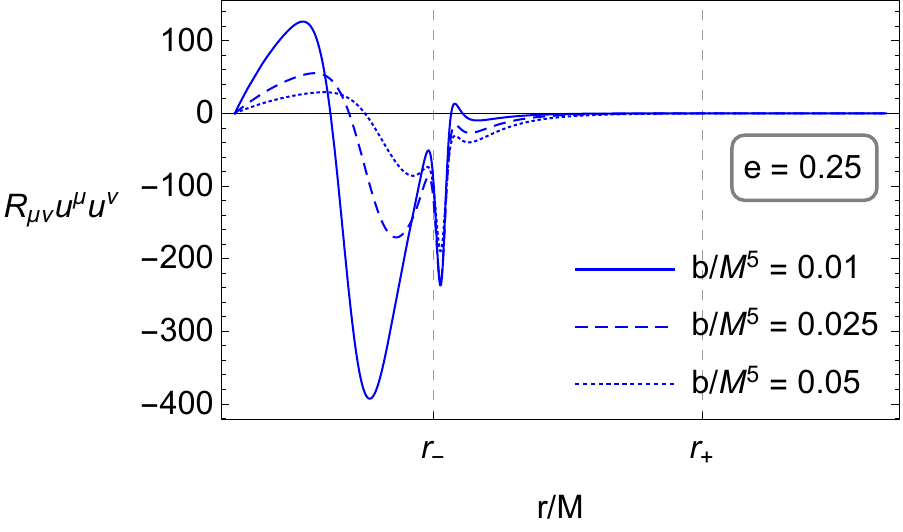}}
    \caption{Contraction of the Ricci tensor with the tangent vector of a particular timelike equatorial trajectory ($L=0,\ E=1$) that falls into a RBH with spin $a/M=0.9$. Each plot is relative to a particular choice of the deviation parameter $e$ and displays curves corresponding to three values of the regularization parameter $b$. The two vertical lines mark the location of the inner and outer horizons.
    }
    \label{fig:SECMplot}
\end{figure}

We then move on to timelike geodesics, whose tangent vector we name $u^\mu$. As before, we choose them to lie on the equatorial plane and to fall into the BH with zero specific angular momentum ($u^\theta=0,\ L=0$); we further choose the radial velocity to be zero at infinity ($E=1$). 
The contraction 
\be
G_{\mu \nu}u^\mu u^\nu
\ee
yields radial profiles that are qualitatively similar to those of \cref{fig:rhoNplot} and for this reason we do not report them here. When this quantity is negative, the weak energy condition is violated.
Finally, we complement the analysis by computing
\be\label{eq:SEC}
R_{\mu \nu} u^\mu u^\nu .
\ee
Assuming the Einstein equations, $R_{\mu \nu} \propto T_{\mu\nu}-(T/2)g_{\mu \nu}$, hence when \cref{eq:SEC} is negative the strong energy condition is violated. 
Some results are reported in \cref{fig:SECMplot}, again for $a/M=0.9$ and a few illustrative choices for $e$ and $b$. As in the null case, these observers measure an effective matter content that is practically zero outside of the outer horizon. Large violations of the strong energy condition are measured inside of the inner horizon. At variance with the null case, now the curves exhibit a second positive bump before reaching zero at $r=0$. 
Similarly to the previous case, increasing the value of $e$ pushes the large negative trough to the right but does not substantially affect its depth, which is instead controlled by $b$; the height of the positive bump increases with $e$. Moreover, for $\abs{e} \lesssim 1$ additional features appear close to the inner horizon. As before, lowering the spin suppresses the magnitude of all these features.

\subsection{Ergosurfaces}

\begin{figure}[ht]
\centering
\includegraphics[width=0.4\textwidth]{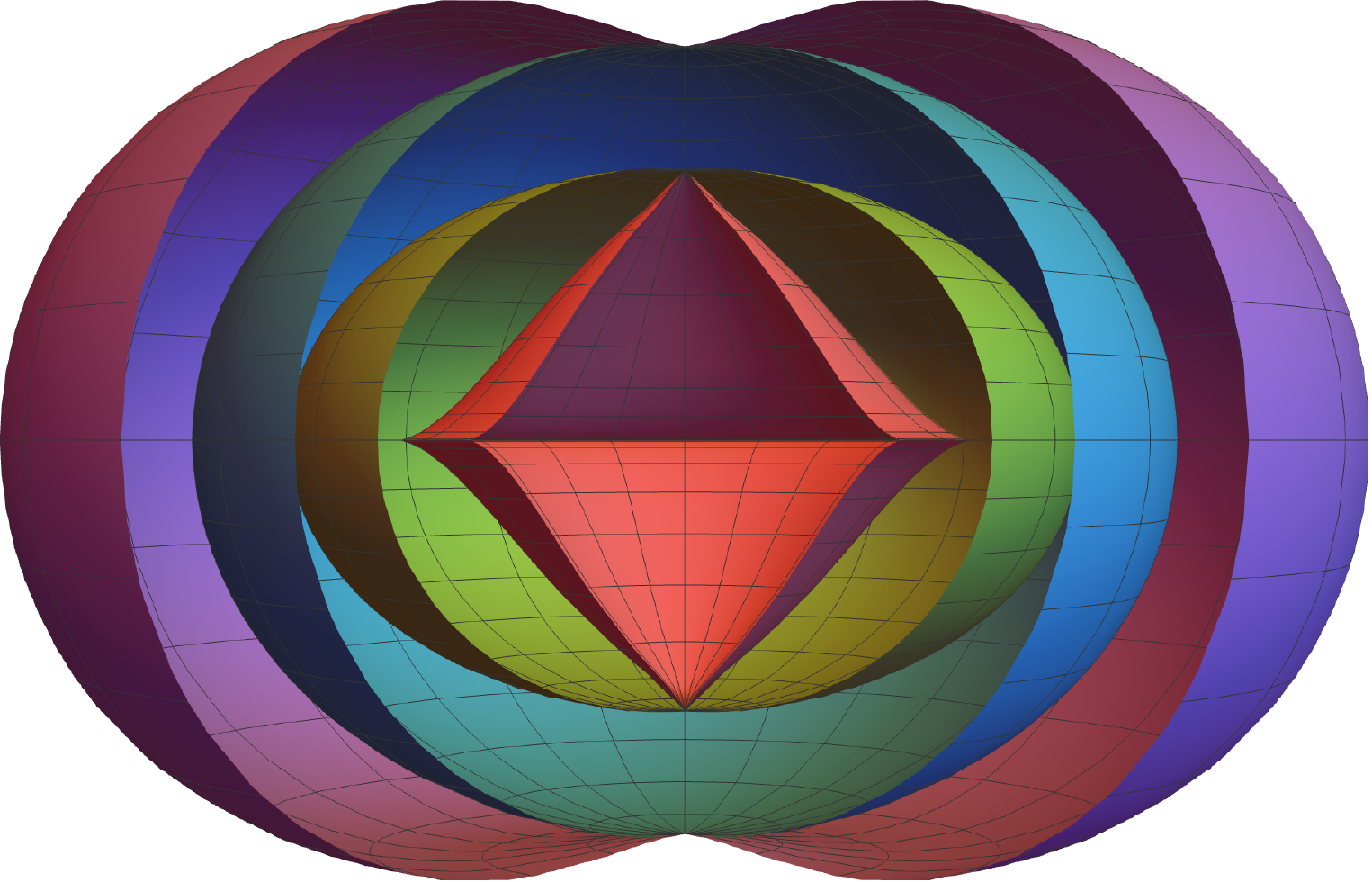}
\caption{Embedding in Euclidean space of the horizons (green and blue surfaces) and ergosurfaces (red and purple surfaces) for $a/M = 0.95$ and $e=1$.}
\label{fig:embed}
\end{figure}

The ergosurfaces are defined by the roots of $g_{tt}=0$, or equivalently of $r^2 - 2m(r)r + a^2\cos^2\theta = 0$, whose solution can be given in closed form. 
Since the result is cumbersome, in \cref{fig:embed} we show the embedding in Euclidean space of the horizons and ergosurfaces for some illustrative choice of the parameters.
The main difference with respect to a Kerr BH is the shape of the inner ergosurface around the poles: values of $e$ closer to the upper and lower bounds in \cref{ebounds} correspond to a more pronounced cuspid around the poles; for values of $e$ closer to the lower bound, the inner horizon and ergosurface move close and eventually touch also at the equator; for values of $e$ closer to the upper bound the horizons move closer as previously said. The conformal factor does not affect the ergosurfaces at all.

Finally, let us notice that, since with our choice $m(r_+)=M$, the textbook expression for the maximal efficiency of the Penrose process~\cite{penrose_extraction_1971, bardeen_rotating_1972, wald_energy_1974, kovetz_efficiency_1975} seems to yield the same result as in Kerr:
\be
\eta_\text{max} = 1-\frac{2m(r_+)}{r_+} = 1-\frac{2M}{M+\sqrt{M^2-a^2}}\, .
\ee
Checking whether this is actually the case would require a more careful analysis of the motion of test particles in our spacetime --- an interesting question which however lies outside the scope of this work.

\subsection{Notable equatorial orbits}

In order to characterize the spacetime and its deviations away from Kerr from a phenomenological point of view, we compute the coordinate location of the light ring (LR) and the ISCO\@.
We focus on the equatorial plane, where the radial motion is governed by the function (cf.\ \cref{rgeod})
\be
\mathcal{R} = E^2r^2(r^2+a^2) - r^2 L^2 + 2m(r) r (aE+L)^2 - \delta\Psi \Delta\,,
\ee
with $\delta=0$ or $1$ for null and timelike geodesics, respectively.
Circular orbits correspond to $\mathcal{R}=\mathcal{R}'=0$ and are stable if $\mathcal{R}''\leq 0$. Since the analytical expressions are not particularly illuminating, the values of $r_\text{LR}$ and $r_\text{ISCO}$ are computed numerically.

\begin{figure*}[ht]
\centering
\includegraphics[width=0.55\textwidth]{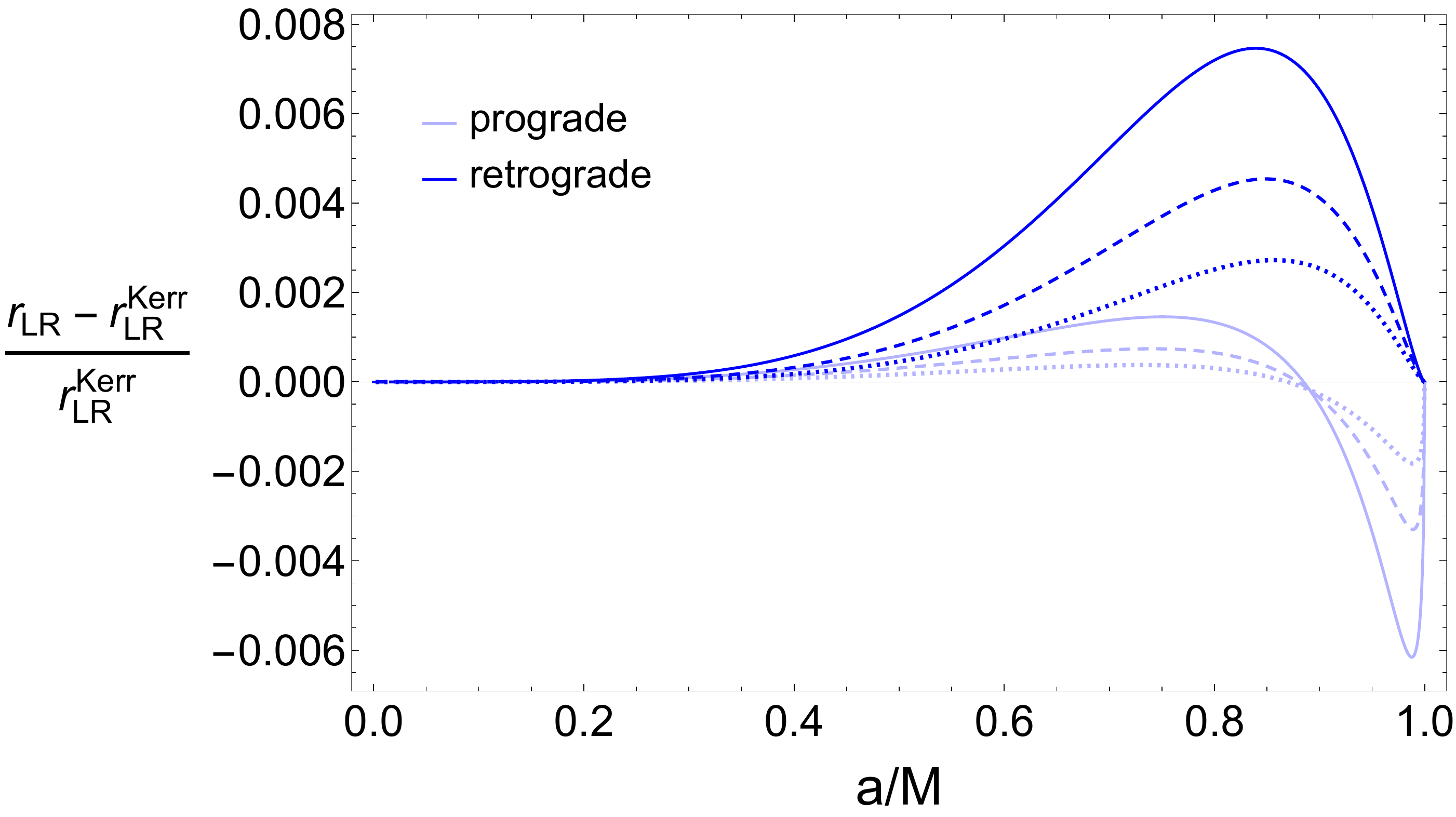}
\caption{Relative difference in the position of the light ring as a function of the spin between Kerr and our spacetime with $e=0.9$ (dotted lines), $e=1$ (dashed lines) and $e=1.1$ (continuous lines). We do not need to specify a value for the regularization parameter $b$ as null geodesics are insensitive to the conformal factor $\Psi$. While we display only values of $e $ near 1, corrections to the light-ring position actually grow very fast with $ e $ and they can be up to order $60\% $ for $e \rightarrow 2$.  Note also that the extrema and sign changes displayed by the curves are ultimately determined by the behavior of the function $m(r)$ (and its derivative), which is not monotonic. }
\label{lightring}
\end{figure*}

\begin{figure*}[ht!]
\centering
\includegraphics[width=0.55\textwidth]{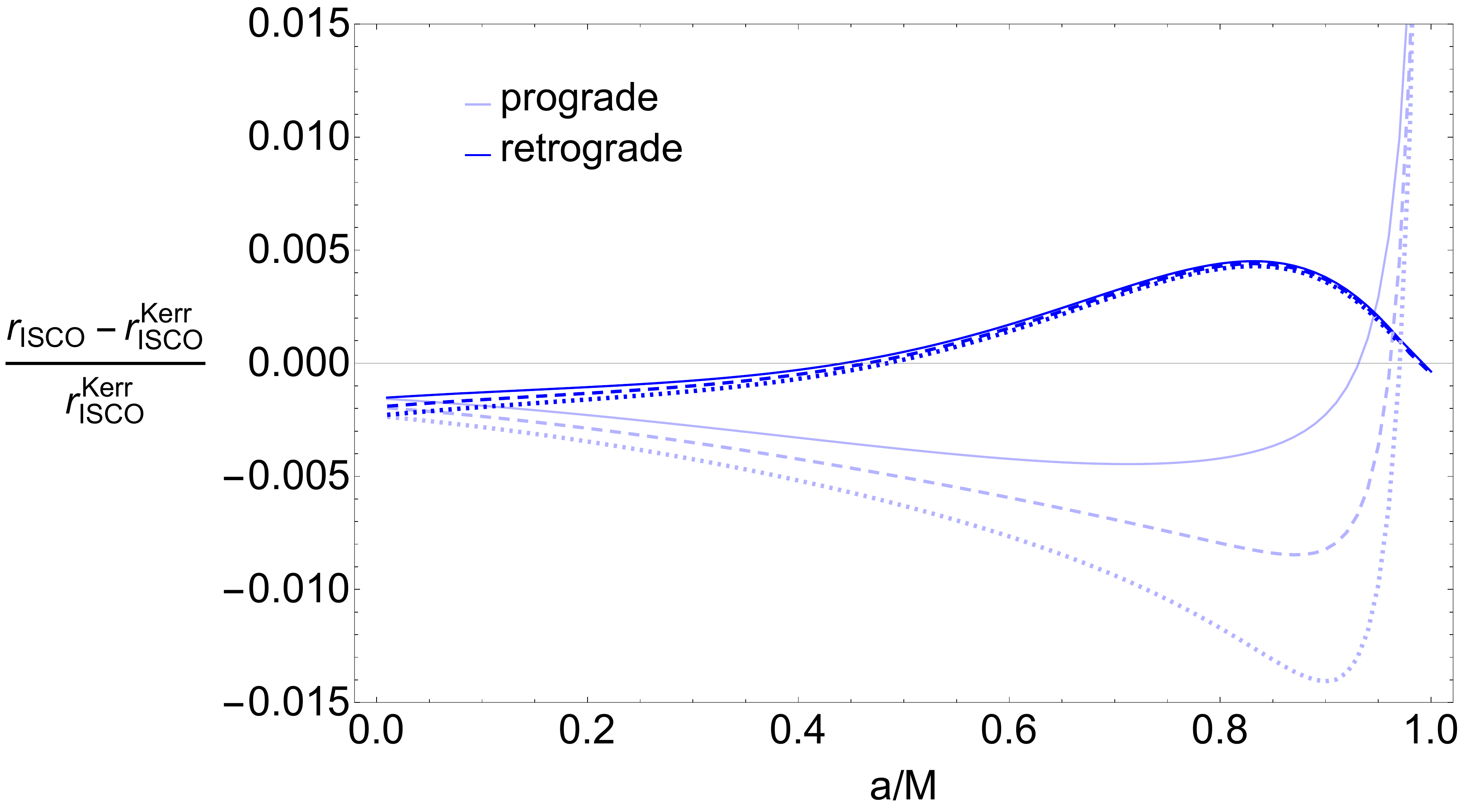}
\caption{Difference in the position of the ISCO between Kerr and our spacetime with $e=1$ and $b/M^5=0.8$ (continuous lines), $b/M^5=1$ (dashed lines) and $b/M^5=1.2$ (dotted lines). The prograde orbit, being in the more internal region of the spacetime where the conformal factor is greater, presents larger deviations, particularly for high spin.}
\label{ISCO}
\end{figure*}

\begin{figure}[ht!]
    \centering
    \subfloat[Prograde orbits.\label{fig:ISCOspanPR}]{\includegraphics[width=0.45\textwidth]{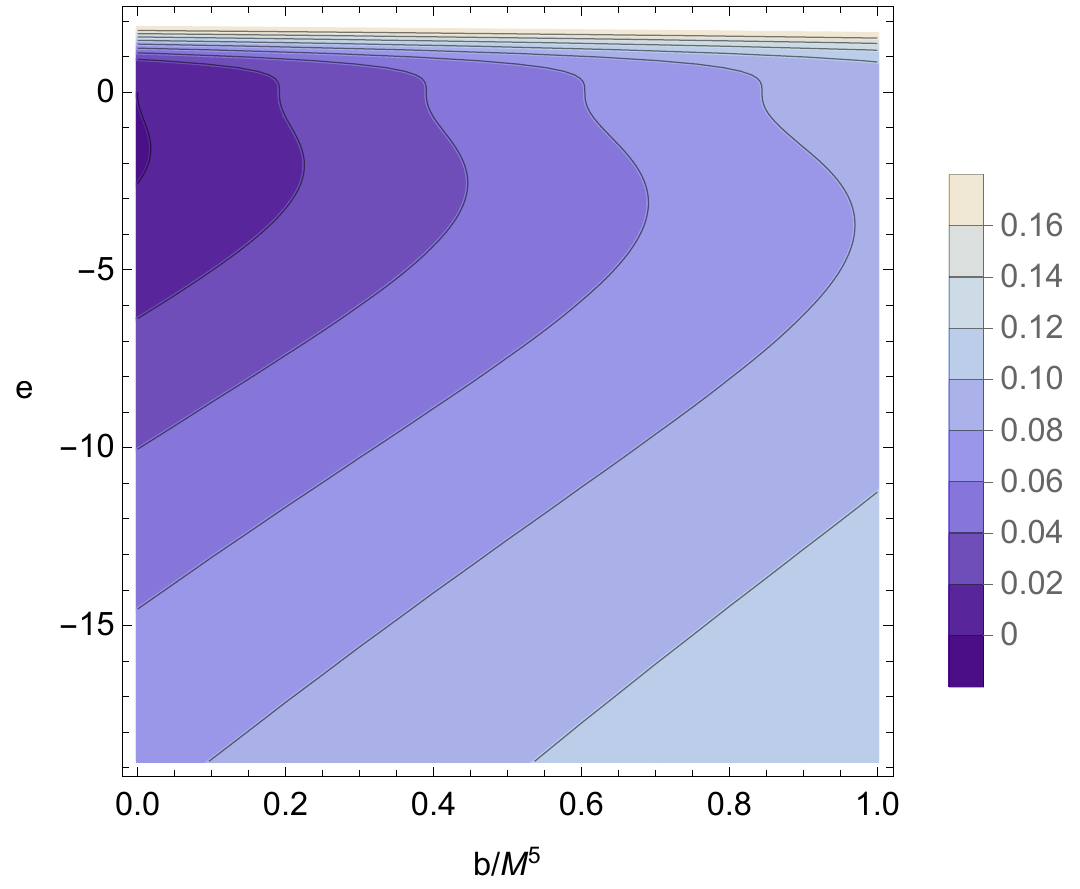}}
    \hfill
    \subfloat[Retrograde orbits.\label{fig:ISCOspanRE}]{\includegraphics[width=0.45\textwidth]{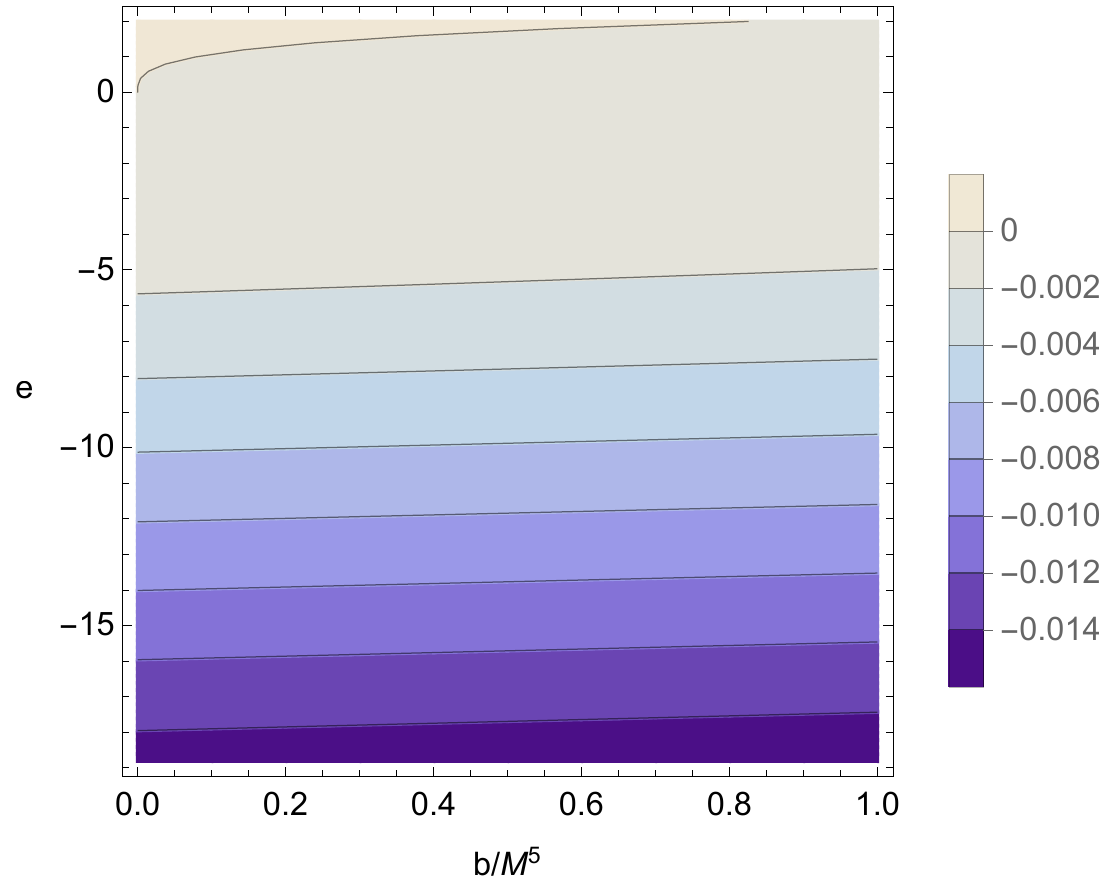}}\\
    \subfloat[Retrograde orbits, zoom to the region $\abs{e}<1$ of \cref{fig:ISCOspanPR}.\label{fig:ISCOspanZOOM}]{\includegraphics[width=0.45\textwidth]{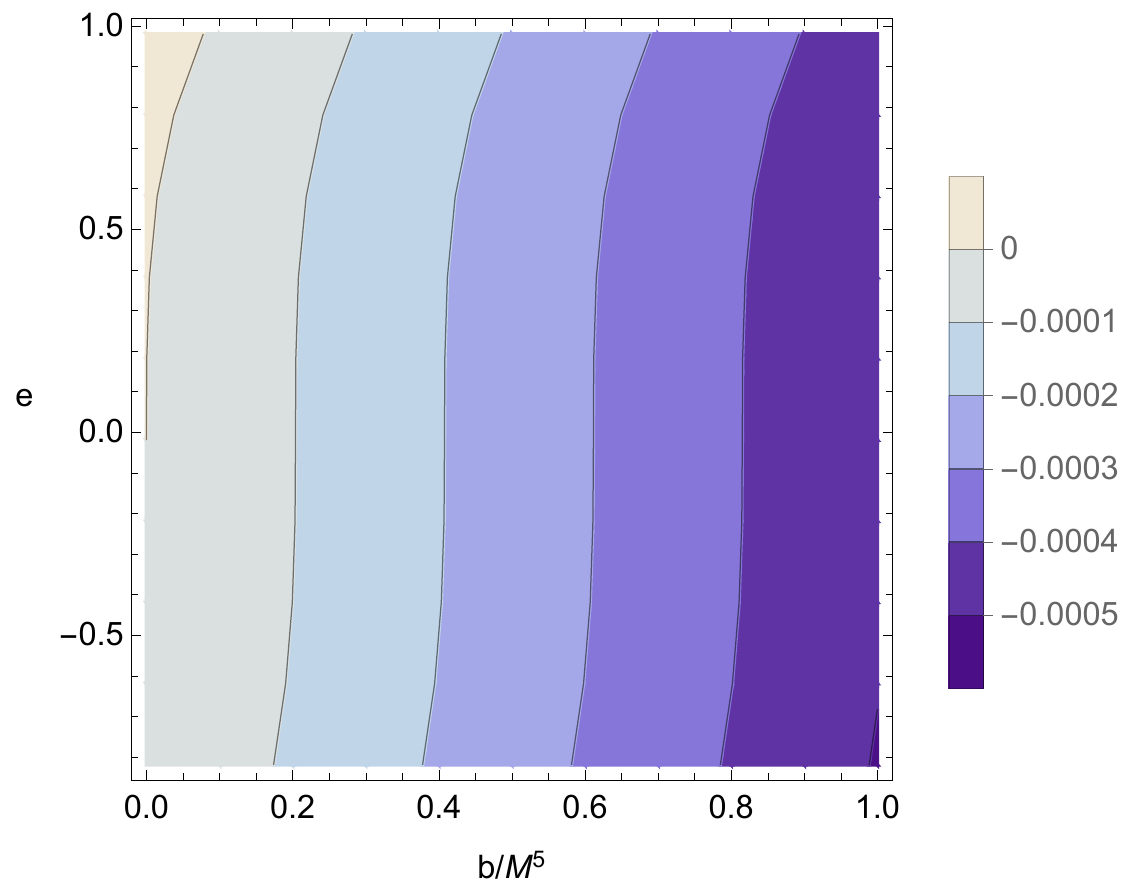}}
    \caption{Fractional deviation of the ISCO from its Kerr analog, computed as $r_\text{ISCO}/r_\text{ISCO}^\text{Kerr}-1\, .$ Spin $a/M=0.998$.}
    \label{fig:ISCOspan}
\end{figure}

The location of the light ring, which is a null geodesic, does not depend on $\Psi$. Its fractional deviation from its Kerr analog is shown in \cref{lightring}, as a function of the spin, for some choices of the parameter $e$. The extrema and sign changes displayed by the curves of \cref{lightring} are ultimately determined by the behavior of the function $m(r)$ (and its derivative), which is not monotonic.
The analogous plot for the ISCO is reported in \cref{ISCO}. 
Contrary to the previous case, $r_\text{ISCO}$ depends on $\Psi$, hence the curves in the figure correspond to specific choices of $b$. In fact, varying the parameter $b$ substantially affects the location of the ISCO, particularly for high spin.  
The peculiar spike associated to prograde orbits and high spin, in particular, can be entirely explained in terms of the behavior of $\Psi$: since, as the spin increases, the prograde ISCO shrinks, $r_\text{ISCO}$ enters deeper into the region where $\Psi$ is markedly different from $r^2$. In order to further explore the parameter space in the high-spin regime, we set $a=0.998M$ (roughly the Thorne limit) and let the parameters vary in the ranges $b\in [0,1]$ and $e \in [-3-3M/\sqrt{M^2-a^2},2]$, thereby producing the contour plots of \cref{fig:ISCOspan}.

Despite the much larger interval spanned by $e$, the gradient of the deviation is dominated by the $b$ component: this is clear for prograde orbits (\cref{fig:ISCOspanPR}), but is also true for retrograde orbits (\cref{fig:ISCOspanRE}) if $e$ is restricted to take reasonably small values as in \cref{fig:ISCOspanZOOM}. Note, however, that even for spins as high as $a=0.998M$, except for rather extreme values of the parameters, the ISCO moves less than a few percent in the prograde case and less than a few per mil in the retrograde case.

\newpage
\section[Conclusions]{Conclusions\label{sec:conclusions}}

In this work we built and studied a new regular alternative to Kerr BHs that is stable under mass inflation. To construct it, we combined the two most common tools for regularization, in a novel way: we used a mass function to construct a degenerate (zero surface gravity) and thus stable inner horizon, and a conformal factor to regularize the singularity. In general, this procedure leads to a family of metrics, depending on the precise choice of the conformal factor and of $m(r)$. 

We decided to focus on a particular form of the conformal factor that accomplishes the regularization in a minimal way and at the same time ensures the non-existence of CTCs and the separability of the equations of motion for test particles. With this choice, the curvature scalars are continuous and tend to zero on the would-be singularity\footnote{With a slightly different choice, however, the limit can also be made non-zero.} thereby solving a long-standing issue that affects many rotating RBHs. The regularization is controlled by a scale that we parametrize in terms of the quantity $b$, with dimensions $[M]^5$. 

We further took an ansatz for $m(r)$ that is again minimal, in a suitable sense, and fixed the coordinate location of the outer horizon so that it coincides with its Kerr analog.
The resulting mass function can be expressed entirely in terms of the coordinate location of the inner horizon, whose deviation from that of Kerr is encoded by the dimensionless quantity $e$.
In the limit $e \rightarrow 0$ we obtain the conformal Kerr metric that, though regular, is characterized by the usual surface gravity at the inner horizon (as it should be given that the surface gravity is conformal invariant) and hence is again unstable under mass inflation. However, it is important to notice that our metric cannot indefinitely deviate from the conformal Kerr one since the deviation parameter $e$ must lie in a specific interval in order for the mass function to be everywhere finite and for the horizons to be well ordered ($0<r_-<r_+$).

Our metric thus depends on a total of four real parameters: the ADM mass $M$, the spin $a$, the regularization parameter $b$ and the deviation parameter $e$. The two additional parameters $b$ and $e$ can be constrained by observations, at least in principle. In particular, $e$ enters at low order in the parametrized post Newtonian expansion of this object gravitational field and thus influences its multipolar structure; moreover, it affects the orbits of massless test particles and therefore shifts the position of the light ring. Finally $e$ and $b$ both affect the motion of massive test particles, with $b$ having the dominant effect on the location of the ISCO (at least when $e$ is taken to vary in reasonably small ranges) especially at high spins.

Let us stress the remarkable relevance of this fact: given that the regularization and Kerr-deviation parameters might be directly related to quantum-gravitational effects, the possibility to constrain them via observations on the exterior geometry of the BH is further evidence that a new window for quantum-gravity phenomenology might be opening via astrophysical observations.

In conclusion, the rotating RBH geometry proposed here is characterized by a wealth of physically desirable features that make it a plausible candidate for the end point of a gravitational collapse to be contrasted with the Kerr geometry to which it can be made arbitrarily close. Constraining the values of the parameters $e$ and $b$ with astrophysical data now becomes an interesting challenge and might represent a new window over testing possible quantum-gravitational scenarios.


\section*{Acknowledgments}
The authors acknowledge funding from the Italian Ministry of Education and Scientific Research (MIUR) under the grant PRIN MIUR 2017-MB8AEZ.


\bibliography{biblio}

\end{document}